\documentclass[aps,floatfix,prb,twocolumn,letterpaper,showpacs]{revtex4-1}
\usepackage[utf8]{inputenc}
\usepackage{graphicx,amsmath}
\usepackage{amssymb}
\usepackage{amsthm}
\usepackage{amsmath}
\usepackage{color}
\usepackage{hyperref}
\usepackage{enumerate}

\hypersetup{
pdfnewwindow=true,      
    colorlinks=true,       
    linkcolor=red,          
    citecolor=blue,        
    filecolor=magenta,      
    urlcolor=cyan 
}
\begin{document}
\title{Thermodynamic and topological phase diagrams of correlated topological insulators}


\author{Damian Zdulski $^{1}$, Krzysztof Byczuk$^{1}$}

\affiliation{$^1$Faculty of Physics, Institute of Theoretical Physics, University of Warsaw,
ul.Pasteura 5, PL-02-093 Warsaw, Poland}
\date{\today}

\begin{abstract}
A definition of topological phases of density matrices is presented. The topological invariants in case of both noninteracting and interacting systems are extended to nonzero temperatures. Influence of electron interactions on topological insulators at finite temperatures is investigated. A correlated topological insulator is described by the Kane-Mele model, which is  extended by the interaction term of the Falicov-Kimball type. Within the Hartree-Fock and the Hubbard I approximations the thermodynamic and topological phase diagrams are determined, where the long-range order is included. The results show that correlation effects lead to a strong suppression of the existence of the nontrivial topological phase. In the homogeneous phase we find a purely-correlation driven phase transition into the topologically trivial Mott insulator.

\end{abstract}

\pacs{03.65.Vf, 71.10.Fd, 71.10.Hf, 71.27.+a}

\maketitle

\section{Introduction}
Recently discovered topological insulators (TIs) have attracted a lot of interest in condensed matter physics.\cite{kanez2,kane3d,kane3d2,zhang,kanerev,kane} 
This novel electronic phase in solids has a band gap in the bulk  but it can conduct an electric  current via gapless edge states which are robust against scattering on impurities or other weak perturbations of  ideal systems.\cite{zhangrev}
 Formation of the  metallic edge states is related to the nontrivial topology of the ground state which originates from a spin-orbit coupling.\cite{geo}
 The TIs are characterized by topological invariants which cannot be continuously  changed unless the single-particle gap is closed.\cite{zhangrev}
  The idea of the TIs  has been extended into symmetry protected topological phases which possess nontrivial topological properties as long as the symmetries are present.\cite{geo} 
  The existence of the topological gapless edge states was experimentally confirmed by  transport measurements\cite{hg,pl,bi,bi2} and an angle-resolved photo-emission spectroscopy\cite{pl,spec1,spec2,spec3} in many materials, such as, e.g., $Bi_{2}Se_{3}$, $Pb_{1-x}Sn_{x}Se$, or $HgTe/CdTe$ quantum wells. 

The existence of the TIs and most of their properties can be understood and described within the noninteracting band theory. Recently, an influence of  electron-electron interactions on the topological insulating states has attracted a growing attention.\cite{corrrev} Theoretical studies are focused mainly on two effects of the strong correlations in topologically nontrivial conditions. The first effect is when a non-on-site interaction yields new topological phases.\cite{corr1} 
The second problem is  how strong correlations change properties of the band topological insulators. 
Namely, interaction can lead to the creation of an ordered phase which through renormalization of the band parameters may compete with the topological state.\cite{corrrev,corr2} 
Moreover, local correlations related to the frequency dependent part of the self-energy can drive a topological phase transition.\cite{corr3} These issues have been stimulated by research on iridium-based materials  $X_{2}IrO_{3}$ (X = Na or Li)  in which both spin-orbit coupling and electron correlations are strong.\cite{top1} Another new class of systems in which strong interactions could play important role are cold atoms in  optical lattices.\cite{opthal}
 Recent experimental progress in this field allows to test Hamiltonians with different types of  hoppings on  lattices, synthetic  spin-orbit couplings and synthetic gauge fields, as well as the interactions, in terms of their topological properties.\cite{opt,spielmann}

Despite of intensive research, the influence of the strong correlations on the TIs at finite temperatures appeared so far only in few studies.\cite{term1,term2,term3,term4}  A major problem was how to extend a idea of the topological insulators to nonzero temperatures. Main progress in this field was made in Ref.~[\onlinecite{topden}] were a concept of topological phases of density matrices was introduced. However, it is still not know how the correlations affect the behavior of the TIs and how to compute topological invariants of interacting systems at nonzero temperatures. 

In this work we investigate the phase diagram of the correlated TIs at finite temperatures. Using the concept of topological phases for density matrices, we introduce methods to compute topological invariants for noninteracting and interacting systems at nonzero temperatures. Then, we consider the Kane-Mele (KM) model without the Rashba spin-orbit coupling term.\cite{kane} To include correlations we extended the KM model by adding an interaction term as in the Falicov-Kimball (FK) model. It describes a short-range local Coulomb interaction between itinerant electrons and localized spinless fermions.\cite{falicov} We employ the Hartree mean-field and Hubbard I approximations to study the phase transitions and obtain the phase diagrams. This approach allows us to obtain semi-analytical results for topological phase transition and examine the influence of the electronic correlations at finite temperatures. Optical lattices offer the possibility of physical realization of this model.\cite{opthal} A similar model has been investigated in Ref.~[\onlinecite{halfal}] within the dynamical mean-field approximation (DMFT). \cite{dmft1,dmft2} However, the long-range ordered phases have not been discussed in general cases.\cite{halfal}

The paper is structured as follows. In Sec.~\ref{sec:s1}. we present the extension of topological invariants to finite temperatures. In Sec.~\ref{sec:s2} we introduce the KM model with the FK type interaction. 
We present the condition for a change of a topological invariant in the system in Sec.~\ref{sec:s3} and in  Sec.~\ref{sec:s3a} we describe the Hartree and Hubbard I approximations.
 In Sec.~\ref{sec:s4} we examine thermodynamic phase diagrams with  topological phases.
In Sec.~\ref{sec:s5} we present our conclusions.

\section{Topological phases at nonzero temperatures}
\label{sec:s1}

\subsection{Topological phases of the density matrices}
\label{sec:s1a}
At a nonzero temperature the system is no longer in  a ground state and there is a general problem how to define and interpret topological invariants. When the electron correlations are absent, the topological phases of insulators at zero temperature can be defined as homotopy equivalence classes of mappings ${\bf k} \mapsto H({\bf k})$, from Brillouin zone (BZ) torus $T^{d}$ to manifold of the gapped Bloch Hamiltonians (BHs).\cite{geo,topclass}  For symmetry protected TIs, points of the manifold represent preserving the symmetry gapped BHs. 

At a nonzero temperature a quantum system is described by a density matrix $\rho$ of a mixed state.  For translational invariant systems without electron correlations the density matrix takes the form $\rho=\prod_{{\bf k}\in BZ}\rho({\bf k})$, where $\rho({\bf k})$ is called the Bloch density matrix (BDM). In order to extend the notion of topological invariants to nonzero temperatures and non-equilibrium systems, topological phases for the BDMs has been defined by using an ensemble of nonorthonormalized pure states, c.f., Ref.~[\onlinecite{topden}].

We express this definition in the equivalent but more convenient for our purposes form, which is analogical as for the BHs, i.e., the homotopy equivalence relation of the BDMs with a constrain that the spectral gap between the valence band eigenvalues $p_{v}({\bf k})$  of $\rho({\bf k})$ and the conduction band eigenvalues $p_{c}({\bf k})$ cannot be closed and protecting symmetries must be preserved during a homotopy transformation. Formally, 

\newtheorem{topdef}{Definition}
\begin{topdef}
\phantomsection
\label{l1}
let $\rho({\bf k})$ and $q({\bf k})$ be Bloch density matrices of size N of a d-dimensional system. Additionally, let corresponding to them density matrices be invariant under a symmetry group $G$. They are said to be topologically equivalent, written as $\rho({\bf k}) \sim q({\bf k})$, if there exist a continuous map $F:T^{d}\times[0,1]\mapsto D_{N}(N)$, such that:

\begin{enumerate}[(i)]
\item $\forall_{{\bf k}}\, \{F({\bf k},0)=\rho({\bf k})\wedge F({\bf k},1)=q({\bf k})\}$,
\item $\forall_{t\in [0,1]}\; \forall_{{\bf k}}\; \{p_{v}({\bf k},t)-p_{c}({\bf k},t)>0\}$,
\item $\forall_{g\in G} \forall_{t\in [0,1]}\; [F(t),U(g)]=0$,
\end{enumerate}
\end{topdef}

where $F(t)$ is the density matrix corresponding to the BDM $F({\bf k},t)$. The eigenvalues $p_{v(c)}({\bf k},t)$ are defined as $n\,(n+1)$th largest eigenvalue of the $F({\bf k},t)$, where the number $n$ is determined by a physical parameter describing occupation of states. In case of equilibrium systems it is the chemical potential $\mu$. It is a generalization of the definition for the BHs, where the valence and conduction bands $E_{v(c)}({\bf k})$ can be defined as $n\,(n+1)$th lowest eigenvalue of $H({\bf k})$. The operator $U(g)$ describes a symmetry operation $g\in G$. The space of the BDMs is labeled as $D_{N}(N)\equiv GL(\mathbb{C},N)/U(N)$. It is the manifold of strictly positive matrices of size $N\times N$, because topological properties are insensitive to scaling \cite{topden} and thereby the normalization condition of BDMs can be omitted. 

The Def.~(\ref{l1}) applies to general BDMs and therefore describes the topological phases for both thermal and non-equilibrium states.\cite{noneq1,noneq2} The crucial role in the Def.~(\ref{l1}) plays the spectral constrain $(ii)$. From it results that to change the topological phase it is necessary to close a spectral gap or break a protecting symmetry.  As a consequence of that, if two topologically nonequivalent systems belonging to the same symmetry class are connected, on the interface between them the transformation $F({\bf k},t)$ evolves through region of the space $D_{N}(N)$ where the spectral gap disappears and thereby edge states occur. These states are insensitive to small perturbations preserving the protecting symmetry. Therefore, the spectral constrain $(ii)$ establishes the bulk-boundary correspondence as in the case of BHs. 

Without the spectral constrain $(ii)$ we can always deform by means of a transformation $F({\bf k},t)$ all BDMs to the maximally mixed one and in that case there exists only the topologically trivial phase. The spectral constrain is not present when all possible states all fully occupied. This entails that sum of topological invariants of all bands must be zero. 

\subsection{Topological invariants at finite temperatures}
\label{sec:s1b}

We show that the Def.~(\ref{l1}) provides for every finite temperature the same classification of the topological phases and formulas for the topological invariants as for BHs. 

\subsubsection{Noninteracting systems}
\label{sec:s1ba}
In equilibrium state the BDM takes the form

\begin{equation}
\label{eqn:e01}
\rho({\bf k})=\frac{e^{-\beta H({\bf k})}}{Z}
\end{equation}
where $Z^{M}=\operatorname{Tr}(e^{-\beta H})$, $H$ is full Hamiltonian, $\beta$ is the inverse of the temperature $T$ and $M$ is number of unit cells. Hence, for every $T$, Eq.~(\ref{eqn:e01}) establishes one-to-one correspondence 
 
\begin{equation}
\label{eqn:e02}
\varphi:H({\bf k})\mapsto\rho({\bf k})
\end{equation}
preserving the protecting symmetries and the spectral gap, i.e., $p_{c}({\bf k})=p_{v}({\bf k})$ if and only if $E_{c}({\bf k})=E_{v}({\bf k})$. As a consequence of that, if two insulators belong to the same topological class of BHs, they must be in the same topological phase of BDMs and thereby $\varphi$ is isomorphism of equivalence classes of topological insulators. Therefore, possible topological phases of finite temperature BDMs have the same classification as in the periodic table of topological insulators and superconductors\cite{kanerev} and topological invariants are computed by substituting

\begin{equation}
\label{eqn:e03}
\varphi^{-1}(\rho({\bf k}))=-\frac{1}{\beta}\ln(Z\rho({\bf k}))
\end{equation}
into the place of $H({\bf k})$ in the standard formulas for the topological invariants.\cite{kanerev,zhangrev} In the limit $T\rightarrow\infty$ this theorem is not valid because $\varphi$ is not injective function, i.e., does not  preserve distinctness of elements. In that case, the system is always described by maximally mixed density of states and thereby it is in the topologically trivial state. 

\subsubsection{Interacting systems}
\label{sec:s1bb}
There is not known a general classification of topological insulators for interacting systems. Their topological invariants are constructed as an analytical continuation of the formulas for noninteracting systems.\cite{geo,intcon} For interacting systems the Hamiltonian cannot be expressed as a single particle operator. Therefore, in order to perform the analytical continuation of the noninteracting topological invariants, they must be formulated in terms of a single particle Green's functions $G$,\cite{intcon} i.e.,
\begin{equation}
\label{eqn:e04}
G^{-1}(\omega,{\bf k})=\omega I-H({\bf k})
\end{equation} 
where $\omega$ is the frequency and $I$ is unity operator. This generalization of topological invariants describes properly topological phases until the interacting system is continuously connected in the space of parameters of the model to some noninteracting system without closing the spectral gap. Hence, this description cannot be applied to fractional topological insulators.

\begin{figure}[th]
\includegraphics[width=0.99\linewidth]{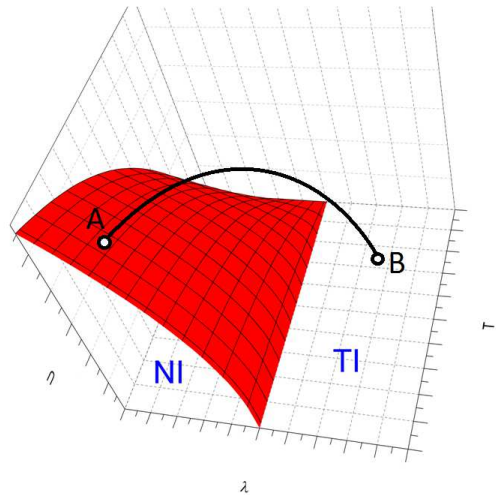}
\caption{Exemplary topological phase diagram in space of interaction $U$, temperature T and parameter $\lambda$. The red surface is the phase boundary separating the normal insulator (NI) and topological insulator (TI). The curve $AB$ connects adiabatically two points.}
\label{fig:f0}
\end{figure}

With additional assumption that $T$ is treated on the same footing as parameters of the model, the above construction can be generalized to the density matrices. This entails that the interacting system can be adiabatically connected to a noninteracting one with a different temperature, what illustrates the line $ AB$ in the Fig.~\ref{fig:f0}.  

As it was discussed in the Sec.~\ref{sec:s1ba} all formulas for the noninteracting topological invariants are identical for BHs and BDMs due to isomorphism from  Eq. (\ref{eqn:e02}). Therefore, by substituting Eq. (\ref{eqn:e03}) into Eq. (\ref{eqn:e04}) in place of $H({\bf k})$  topological invariants expressed by $G(\omega,{\bf k})$ can be calculated for the BDMs. The analytical continuation to interacting systems does not break this correspondence and thereby the same statements hold for the topological invariants of the interacting systems. In particular, also the concept of topological Hamiltonian \cite{topham1,topham2} remains unchanged for the density matrices.

\subsection{Interpretation of topological invariants at finite temperatures}
\label{sec:s1c}
In recently published papers Ref.~[\onlinecite{inv1}] and [\onlinecite {inv2}] different extension of the concept of topological invariants to finite temperatures is proposed. Those topological invariants do not match with these derived here from the Def.~(\ref{l1}). In contrast to our proposal, those topological invariants depend explicitly on a temperature for noninteracting systems. Additionally, it is shown in Ref.~[\onlinecite{topden}] that they are not unique.  Another advantage of the Def.~(\ref{l1}) is that due to the bulk-boundary correspondence the spectral constrain $(ii)$  links our topological invariants to the physical observable, which is the number of edge states. 

Finally we note that at nonzero temperatures, the Hall conductivity is not quantized due to a thermal activation of states from a conduction band. Therefore, the topological invariants cannot be measured directly as response functions. However, at sufficiently low temperatures the contribution from the edge states should dominate. On the other hand, the edge states are experimentally accessible by angle-resolved photo-emission spectroscopy even at the room temperature.\cite{pl} 

\section{Model}
\label{sec:s2}
As discussed in the Sec.~\ref{sec:s1}, topological invariants for noninteracting systems at finite and zero temperature are the same, when parameters of the model are not temperature depended. Situation is more complex in case of correlated systems because the temperature change of the order parameter affects the self-energy of the electrons and hence the topological properties of the system. To investigate this effect we use the KM model.\cite{kane}
It is represented by the Hamiltonian which eigenstates exhibit  the quantum spin Hall effect.\cite{hirsch} Formally, it is constructed from two copies of the Haldane model\cite{haldane} with opposite signs of the Peierls phase for the electrons with the up and down spins, respectively, moving on a honeycomb lattice. We extend this model by adding localized spinless fermions which locally interact with the itinerant electrons as in the FK model.\cite{falicov} 
The full Hamiltonian is
\begin{equation}
\begin{split}
\label{eqn:e1}
H=&-t_1\sum_{\langle i,j \rangle\sigma}c^{\dag}_{i\sigma}c_{j\sigma}-t_{2}\sum_{\ll i,j \gg\sigma \sigma '}e^{i\phi_{ij}}c^{\dag}_{i\sigma}\tau^{z}_{\sigma\sigma '}c_{j\sigma '}\\
&+U\sum_{i\sigma}(c^{\dag}_{i\sigma}c_{i\sigma}-\frac{1}{2})(f_{i}^{\dag}f_{i}-\frac{1}{2}),
\end{split}
\end{equation}
where $c_{i\sigma}$ ($c_{i\sigma}^{\dagger}$) is the annihilation (creation) operator for the itinerant electron with the spin $\sigma=\uparrow$ ($\downarrow$), and $f_i$ ($f_i^{\dagger}$) is  the annihilation (creation) operator for the localized spinless  fermion at the lattice site $i$.  The parameter $t_1>0$ is the nearest-neighbor hopping whereas $t_{2}>0$ is the amplitude of the next-nearest-neighbor hopping. We use $t_1=1$ which sets the energy units. The Peierls phase is $\phi_{ij}=\pm \frac{\pi}{2}$ for the  next-nearest-neighbor hopping in the clockwise and anticlockwise direction, respectively. The third Pauli matrix is $\tau^{z}_{\sigma\sigma'}$ and it describes the change of the sign in the Peierls phase for the electrons with opposite spins. The strength of the local FK interaction is given by $U$.
 In this work we consider a  half-filling, i.e. one particle per unit cell for both kinds of fermions.
 Then the chemical potentials for the  itinerant ($\mu_c$)  and localized ($\mu_f$) particles are $\mu_c=\mu_f=0$.
 The Hamiltonian is invariant with respect to: the lattice inversion, the spin rotation, the time reversal symmetry, and the particle-hole transformation. Due to particle-hole symmetry the system does not possess  an indirect band gap for any parameters. 

The localized particles in the FK model are thermodynamically coupled to the itinerant electrons and their spatial distributions is  determined by a minimum of  the thermodynamic potential of the system.  
On a  bipartite lattice at half-filling  the solution of the Kane-Mele Falicov-Kimball (KMFK) model Eq.(\ref{eqn:e1}) with $t_2=0$ possesses a long-range order at low temperatures,\cite{lieb} in which the localized fermions form a checkerboard pattern and the itinerant electrons form a charge density wave (CDW)~.
Then the inversion symmetry of the ground state  is broken. The transition between the CDW phase and the homogeneous phase is continuous.  The order parameter is $d\equiv (n_{B}-n_{A})$, where $n_{\alpha}$ is the average number of localized particles per unit cell in the sublattice $\alpha ={\rm A}$ or $\rm B$. 
Increasing the hopping amplitude $t_2$ can lead to a change of the ground state.\cite{wojt} We estimate that the CDW phase is stable  for $t_{2} \le 0.5 t$ for an arbitrary U (see Appendix \ref{sec:a1}). Therefore, we only present results for this regime of the $t_2$ hopping parameter.

\section{Topological phase transition}
\label{sec:s3}

\subsection{General conditions}

Topological phases are quantified by topological invariants. For the model (\ref{eqn:e1}) its topological properties are fully determined by the topological invariant equivalent to the Chern number $C$ for a single spin subsystem.\cite{chern}  In the noninteracting limit and $t_{2}>0$ the system is in a topologically nontrivial phase  with two edge states for open boundary conditions.\cite{kane}

In order to study topology of the solution for the model (\ref{eqn:e1}) with finite  $U$ we introduce the one-particle Green's function $G_{\sigma}(\omega, {\bf k};d)$, which depends implicitly on the order parameter $d$ for the CDW phase. 
The Green's function obeys the Dyson equation
\begin{equation}
\label{eqn:e2}
G^{-1}_{\sigma}(\omega,{\bf k};d)=(\omega+\mu_c)I-H_{0\sigma}({\bf k})-\Sigma_{\sigma}(\omega,{\bf k};d),
\end{equation}
where $$\Sigma_{\sigma}(\omega,{\bf k};d)=\left(
\begin{array} {cc}
\Sigma_{\sigma A}(\omega,{\bf k};d) & 0 \\
 0 & \Sigma_{\sigma B}(\omega,{\bf k};d)
\end{array}\right) $$
 is the diagonal self-energy matrix of the interacting system and $H_{0\sigma }({\bf k})$ is a matrix representation of the noninteracting Hamiltonian. From the theorem discussed in the Sec.~\ref{sec:s1b} results that, topological invariant at every finite temperature for our interacting system is determined by using the standard method\cite{kanerev} applied to the topological Hamiltonian defined by the  noninteracting Hamiltonian matrix and the self-energy at $\omega=0$,\cite{topham1,topham2} i.e. 
\begin{equation}
\label{eqn:e4}
H^{\rm top}_{\sigma}({\bf k};d)=H_{0\sigma }({\bf k})+\Sigma_{\sigma}(\omega=0,{\bf k};d).
\end{equation}
Introducing symmetric and antisymmetric combinations of the self-energies 
$$\Sigma_{\sigma \rm S/R}({\bf k};d)=\frac{\Sigma_{\sigma A}(0,{\bf k};d)\pm\Sigma_{\sigma B}(0,{\bf k};d)}{2},$$
the topological Hamiltonian for the spin $\sigma$ particles reads
\begin{equation}
\label{eqn:e5}
H^{\rm top}_{\sigma}({\bf k};d)=H_{0\sigma}({\bf k})+ \Sigma_{\sigma \rm S}(0,{\bf k};d)I+\Sigma_{\sigma \rm R}(0,{\bf k};d)\tau^{z}.
\end{equation}
The last term breaks explicitly the inversion symmetry. In rest of the paper we use local approximations in which the self-energy does not depend on the wave vector $\bf k$, i.e.  $\Sigma_{\sigma \rm S/R}({\bf k}; d)= \Sigma_{\sigma \rm S/R}(d)$ is only a function of the order parameter $d$. 

Occurrence of the topologically nontrivial phase is connected with the fact that we cannot choose a one global gauge for eigenvectors of the Hamiltonian matrix (\ref{eqn:e5}) that is continuous and single valued over whole BZ.\cite{booktop} This yields the condition 
\begin{equation}
\label{eqn:e6}
|\Sigma_{R}(d)|>3\sqrt{3}t_{2}  
\end{equation}
for the  transition from the topological phase  to the trivial phase (see Appendix \ref{sec:a2}). We note that this is a general condition for existence of nontrivial topology for any local approximation.

\subsection{Considered approximations}
\label{sec:s3a}

\subsubsection{Hartree approximation}

Within  the Hartree approximation the self-energy (\ref{eqn:e2}) of the itinerant particles is given by\cite{bookgreen}
\begin{equation}
\label{eqn:e7}
\Sigma_{\sigma}(\omega,{\bf k};d)=\left(
\begin{array} {cc}
U\,n_{A} & 0 \\
 0 & U\,n_{B}
\end{array}
\right).
\end{equation} 
Hence, 
\begin{equation}
\label{eqn:e7a}
\Sigma_{\sigma R}(d) = -\frac{Ud}{2}
\end{equation}
and from Eq. (\ref{eqn:e6}), the condition for the change of the topological invariant takes the form
\begin{equation}
\label{eqn:e8}
U>U_c^{\rm H}=\frac{6\sqrt{3}t_{2}}{|d|}.
\end{equation} 
The Hartree approximation corresponds to the first order perturbation expansion with respect to small $U$.

\subsubsection{Hubbard I approximation}

Within the Hubbard I approximation the self-energy  (\ref{eqn:e2}) of the itinerant particles is given by\cite{bookgreen}
\begin{equation}
\label{eqn:e9}
\setlength{\arraycolsep}{-2pt}
\Sigma_{\sigma}(\omega,{\bf k};d)=\left(
\begin{array} {cc}
U\,n_{A}+\frac{U^{2}n_{A}n_{B}}{\omega+U(\frac{1}{2}-n_{B})} & 0 \\
 0 & U\,n_{B}+\frac{U^{2}n_{B}n_{A}}{\omega+U(\frac{1}{2}-n_{A})} 
\end{array}
\right).
\end{equation} 
\\
Hence, 
\begin{equation}
\label{eqn:e9a}
\Sigma_{\sigma \rm R}(d)=-\frac{Ud}{2}-\frac{U(1-d^{2})}{2d}=-\frac{U}{2d}.
\end{equation}
There are two contributions to Eq.  (\ref{eqn:e9a}): (i) the Hartree term, which takes into account a formation of the long-range order; and (ii) the term that takes into account effects of local electron correlations. Substituting $\Sigma_{\sigma \rm R}$ into the Eq. (\ref{eqn:e6}), the condition for the change of the topological invariant takes the form
\begin{equation}
\label{eqn:e10}
U>U_c^{\rm HI}=6\sqrt{3}t_2|d|.
\end{equation} 
The Hubbard I approximation becomes exact when we neglect hopping $t_1$ and $t_2$ amplitudes. 

\subsection{Topological phase diagrams}

\begin{figure}[th]
\includegraphics[width=0.99\linewidth]{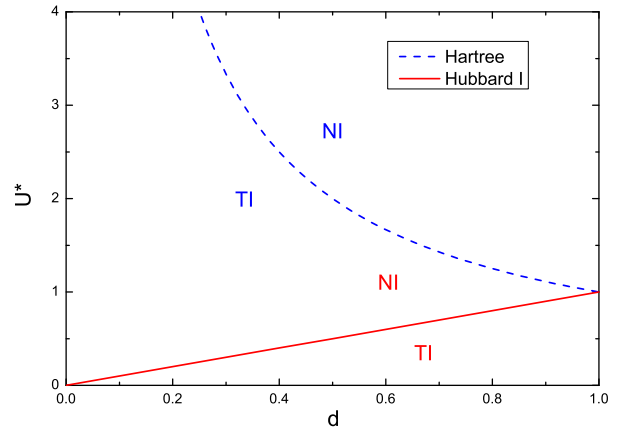}
\caption{Topological phase diagram of KMFK model within the Hartree (blue line) and  Hubbard I (red line) approximations in the space of ($d$, $U^{*}$) parameters, where: $U^{*}=U/6\sqrt{3}t_{2}$, TI - topological insulator, NI - normal insulator.}
\label{fig:f1}
\end{figure}

\begin{figure}[th]
\includegraphics[width=0.99\linewidth]{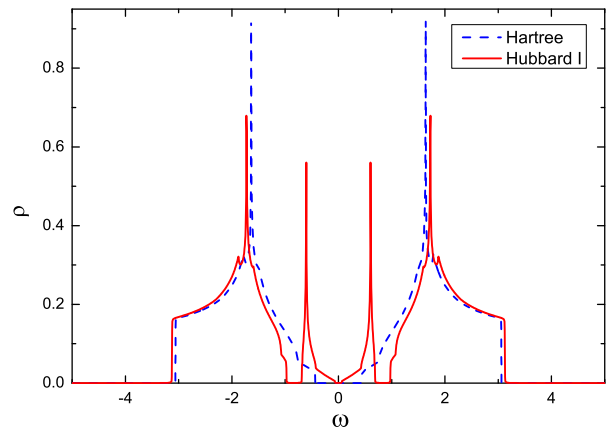}
\caption{Density of states of KMFK model within the Hartree and Hubbard I approximations.  We use: $t_{2}=0.2$, $U=1.5$, and $d=0.8$. }
\label{fig:f2}
\end{figure}

In  both Eqs.~(\ref{eqn:e8}) and (\ref{eqn:e10}) the critical interaction strength $U_c$ depends linearly on $t_{2}$. For $|d|=1$ the correlation term in (\ref{eqn:e9a}) vanishes and $U_c^{\rm HI}=U_c^{\rm H}$. 
For $|d|<1$ the critical interaction $U_c^{\rm H}$ increases with decreasing $|d|$ whereas $U_c^{\rm HI}$ vanishes linearly with decreasing $|d|$, Cf.~Fig.~\ref{fig:f1}. Since the electronic correlations are included within the Hubbard I approximation we conclude that they reduce strongly the topologically nontrivial solution. In details, this is due to formation of additional band states  inside the Hartree gap, as is shown in  Fig.~\ref{fig:f2}, because of the frequency dependence of the self-energy (\ref{eqn:e9}). This results  that the energy gap is closed for smaller  $U$ in the comparison to the Hartree approximation case and thereby the topological invariant is changed.

In a homogeneous phase with $d=0$ the noninteracting model has a topological insulating phase. For any finite interaction  $U>0$ the Hubbard I solution gives a  topologically trivial Mott insulator. It means that  $U=0$ is a singular point in this approximation.  The transition from the topological insulator to the normal Mott insulator occurs  without closing the bad gap and it is only driven by local correlations. We find that  in the Hubbard I approximation $\Sigma_{R}(d\rightarrow0)=\infty$ for any $U>0$. However, the analytical condition Eq. (\ref{eqn:e10}) remains well-defined if we take $d \rightarrow 0 $ limit at the end. Then, we do not have to use the concept of the frequency domain winding number \cite{freq} to compute the topological invariant in the homogeneous phase.

\subsection{More advanced approximations}
In Fig.~\ref{fig:f1} we see that the Hartree and Hubbard I solutions diverge from each other when the homogeneous phase is approached. 
Mathematically, it is due to the fact that when the long-range order vanishes the correlation part of the Hubbard I self-energy  becomes singular.
Physically, it reflects the fact that these two approximations are valid in different limits, as discussed earlier. 
To resolve that problem one has go beyond the perturbation theory because  one cannot get  the Mott insulator by any local non-renormalized perturbation expansion (NRPE) in $U$.\cite{Lutt} 
Indeed, within NRPE we find the following self-energy for the homogeneous phase 
$$\Sigma_{\sigma}(\omega,{\bf k};d=0)=\frac{U}{2}+\frac{U/4}{\omega-\Delta_{0}(\omega)},$$ 
where the hybridization function  $\Delta_{0}(\omega)=\omega-G^{-1}_{0}(\omega)$ and $G_{0}(\omega)$ is a noninteracting Green function at a single lattice site and is given by  the Hilbert transform of a noninteracting density of states $G_{0}(\omega)=\int \rho_{0}(\omega)/(\omega-\epsilon)$. 
From the particle-hole symmetry $G_{0}(0)=-i\pi\rho(0)$, where $\rho(0)=0$ in our system. 
Hence, $\Delta_{0}(\omega\rightarrow0)=-i\infty$ and thereby at $\omega=0$ the self-energy reduces to the Hartree self-energy.

The most advanced renormalized local approximation is the DMFT.\cite{dmft1,dmft2}  Our preliminary DMFT results\cite{future} show that up to the critical  interaction strength $U_{c1}$ the  hybridization function at $\omega=0$ is the same as within the Hartree approximation and the system is a topological insulator. Then, the metal-insulator transition occurs and the hybridization function has finite imaginary part. At the interaction strength $U_{c2}> U_{c1}$ the gap is opened again and the systems becomes topologically trivial Mott insulator with the hybridization function at $\omega=0$ equal to zero as within  the Hubbard I approximation. Hence, the Hartree solution can be interpreted as an upper bond and the Hubbard I as a lower bond on the value of $U_{c1(2)}$, respectively. This result is consistent with a work,\cite{halfal} in which the DMFT equations where solved in the homogeneous phase.

\section{Thermodynamic phase diagram}
\label{sec:s4}

Stability of the CDW phase is determined by minimization of the free energy with respect to the order parameter  $d$ at a fixed temperature $T$ and given microscopic parameters $U$ and $t_2$. The free energy of the KMFK model is\cite{free}
\begin{equation}
\begin{split}
\label{eqn:e14}
F(d,T)&=- \frac{2}{\beta}\int \rho(\omega;d)\ln(1+e^{-\beta\omega})d\omega\\
&+ \frac{1}{\beta} [n_{A}\ln(n_{A})+n_{B}\ln(n_{B})], 
\end{split}
\end{equation}
where the first term describes a  contribution from the itinerant electrons and is equivalent  to the free energy of the noninteracting fermions but with 
a density of states $\rho(\omega;d)$ modified by the interaction. The second term describes an entropic contribution from the localized particles. 
The inverse of the temperature is denoted by $\beta=1/T$. The temperature dependence of the topological phase transitions is obtained by inserting the optimal value $d(T)$ into corresponding Eqs.~(\ref{eqn:e8})  and (\ref{eqn:e10}). 

\begin{figure}[htp!]
\includegraphics[width=0.99\linewidth]{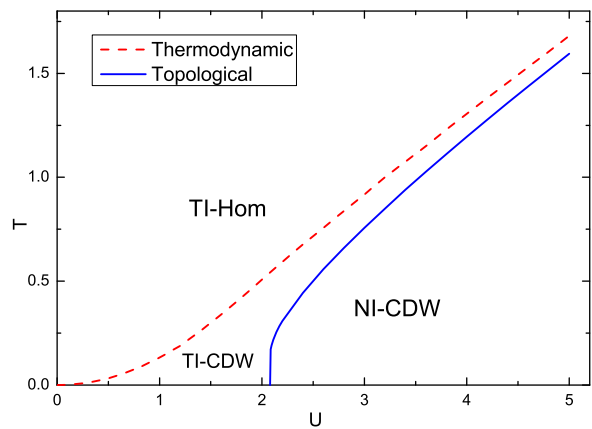}
\caption{Phase diagram of  KMFK model for $t_{2}=0.2$ within the Hartree approximation.}
\label{fig:f3}
\end{figure}

The thermodynamic and topological  phases within the Hartree approximation are displayed in Fig.~\ref{fig:f3} for a selected value of $t_2=0.2$. 
There are three phases: a homogeneous topological insulator (TI-Hom), a topological insulator with the CDW long-range order (TI-CDW), and a normal insulator with the CDW long-range order (NI-CDW). 
The continuous thermodynamic transition between the TI-Hom and the TI-CDW is marked by the red dashed curve. 
The topological  transition between the TI-CDW and the NI-CDW, displayed by the blue solid curve, is associated with closing a gap and a semi-metallic behavior right at the transition line. Interestingly, the coexistence of the CDW long-range order and the topological phase occurs in a finite range of $U$ values for a given temperature.  
This also contrasts to the Haldane model with nearest-neighbor interactions, where the charge-ordered phase is always topologically trivial.\cite{next}
Since at $T=0$ the order parameter $d=1$ for any $U$, we find from Eq.~(\ref{eqn:e8}) that the TI-CDW - NI-CDW transition takes place at $U_c^{\rm H}=6\sqrt{3}t_2$, which agrees with the Fig.~\ref{fig:f3}.
 For other $t_2$ hopping parameters the phase diagrams are qualitatively similar to that presented in Fig.~\ref{fig:f3}.
  However, when $t_2$ decreases the coexistence TI-CDW  regime shrinks and finally disappears at $t_2=0$ because it is $t_2$ that induces the topologically nontrivial phase.

 \begin{figure}[htp!]
\includegraphics[width=0.99\linewidth]{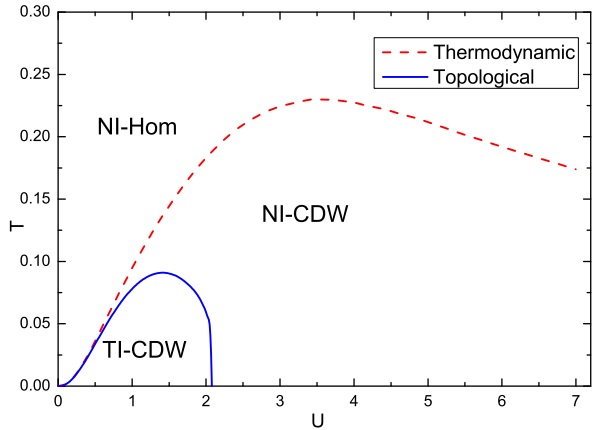}
\caption{Phase diagram of the studied model for $t_{2}=0.2$ within Hubbard I approximation.}
\label{fig:f4}
\end{figure}

The thermodynamic and topological  phases within the Hubbard I approximation are displayed in Fig.~\ref{fig:f4} for the same $t_2$ as in the Hartree approximation above. There are again three but different phases: a homogeneous normal insulator (NI-Hom), TI-CDW, and NI-CDW. The homogeneous topological insulator, seen in Fig.~\ref{fig:f3}, is replaced by the homogeneous normal insulator. The topological phase occurs together with the CDW long-range order and forms a bounded area on the phase diagram with a non-monotonic behavior of the critical temperature. Again this is in contrast with the Haldane model with nearest-neighbor interactions, where the charge-ordered phase is  topologically trivial.\cite{next} We also note that at $T=0$ the phase diagrams Figs.~\ref{fig:f3}~and~\ref{fig:f4} are identical. This is a general feature of the  solutions of the  KMFK model within local approximations to the self-energy. 

\begin{figure}[thp!]
\includegraphics[width=0.99\linewidth]{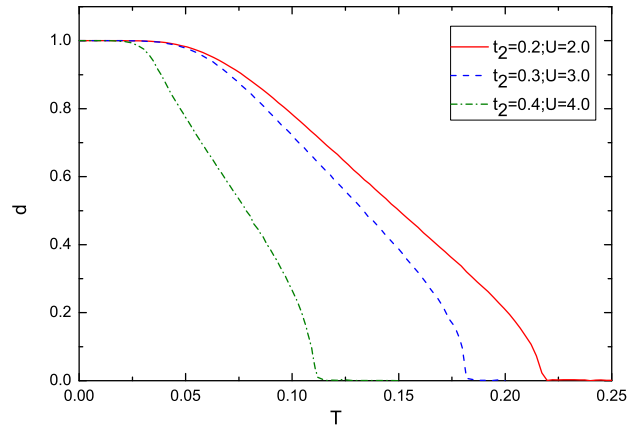}
\caption{Order parameter as a function of the temperature for selected $t_{2}$ and $U$ within the Hubbard I approximation.}
\label{fig:f7}
\end{figure}

The topological transition line in Fig.~\ref{fig:f4}  has two characteristic behaviors: (i) for small U the  thermodynamic and topological lines are almost tangent to each other; and (ii) for U close to $U_c^{\rm HI}=6\sqrt{3}t_2$ the topological line is almost vertical. The course of the topological line results from Eq.~(\ref{eqn:e10}) which relates the critical interaction strength $U_c^{\rm HI}$ and the value of the order parameter $d(T)$ for which topological transition occurs. For temperatures close to the thermodynamic transition temperature $T_c$,  the order parameter takes the form $d(T)\approx \alpha\sqrt{T-T_c(U)}$, where $\alpha$ is a constant. Substituting this function into  Eq.~(\ref{eqn:e10}) the topological transition temperature $T_{\rm top}$ is
\begin{equation}
\label{eqn:e15}
T_{\rm top}\approx T_{c}(U)+\frac{U^{2}}{108\alpha^{2}t_{2}^{2}}.
\end{equation}
Hence, the topological critical temperature tends  to the thermodynamic one, when U tends to zero and its main correction is of the order $O(U^2)$. In the second case the topological line follows the course of the order parameter when its value approaches unity. At low temperatures the order parameter is almost a constant, as is seen in Fig.~\ref{fig:f7}, and this behavior gives rise  to  the vertical slope of the topological transition line in Fig.~\ref{fig:f4}. 

\begin{figure}[thp!]
\includegraphics[width=0.99\linewidth]{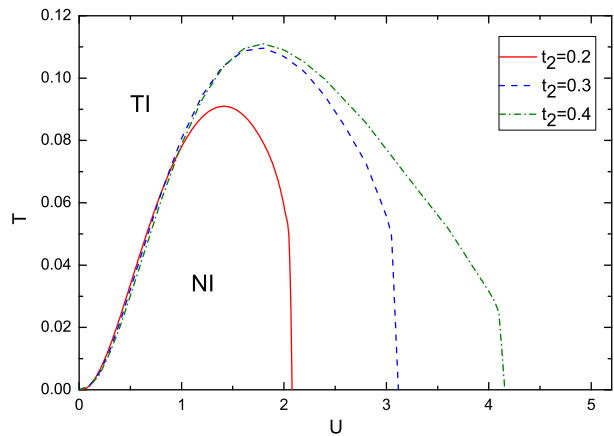}
\caption{Topological phase diagrams of the KMFK model for selected $t_{2}$ within the Hubbard I approximation.}
\label{fig:f6}
\end{figure}

The topological phase diagrams for different $t_2$ are displayed in Fig.~\ref{fig:f6}. When the hopping parameter increases, an area of the topological phase is extended. This originates from the fact that the hopping $t_2$ triggers the topological phases in the KMFK model (\ref{eqn:e1}). The height of the nearly vertical part of the topological line decreases because the constant part of the function $d(T)$ decreases with increasing $t_2$,  as is seen in Fig.~\ref{fig:f7}.

\begin{figure}[htp!]
\includegraphics[width=0.99\linewidth]{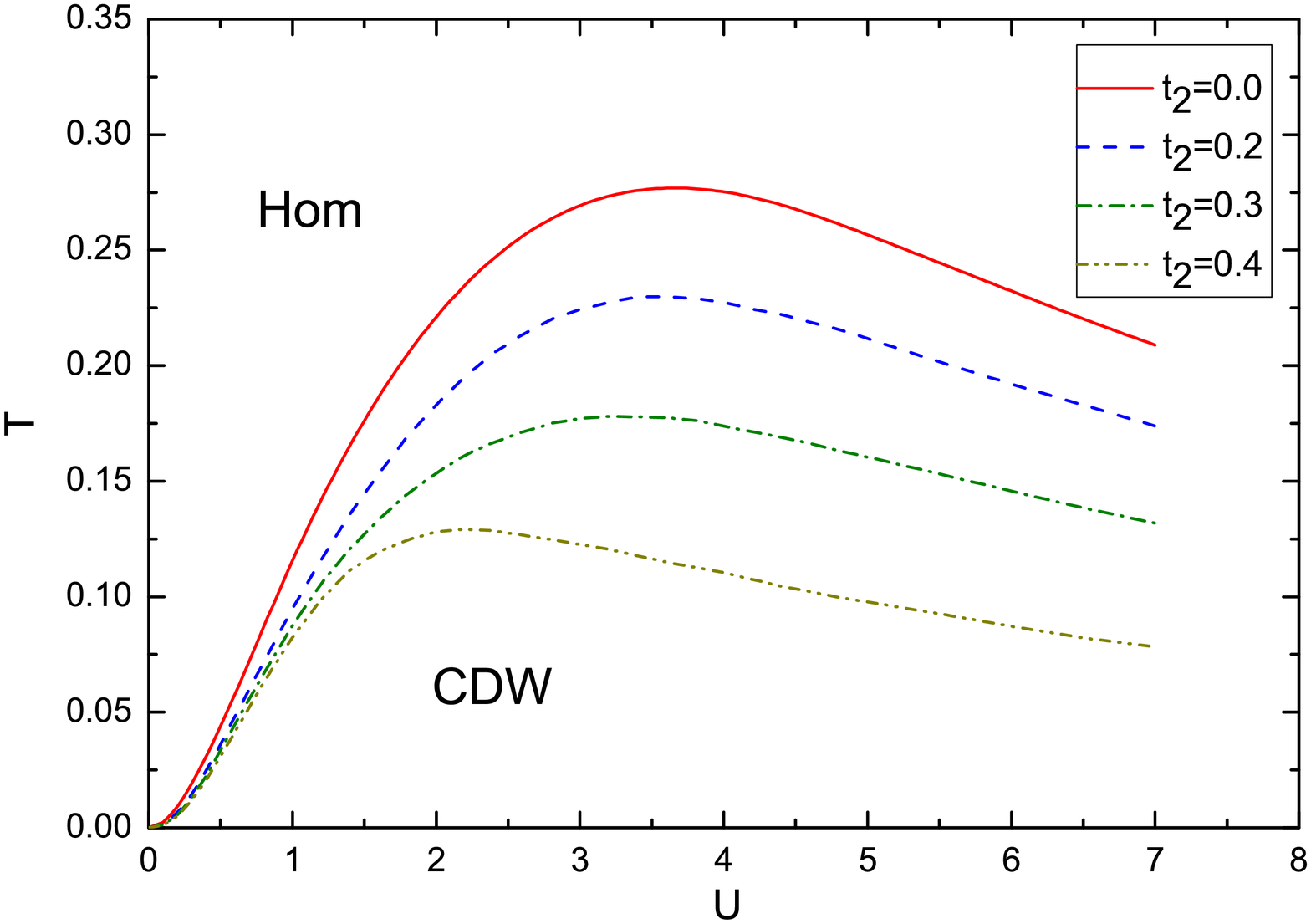}
\caption{Thermodynamic phase diagrams of the KMFK model for selected $t_{2}$ within the Hubbard I approximation.}
\label{fig:f5}
\end{figure}

Comparing  Figs.~\ref{fig:f3}~and~\ref{fig:f4} we see that the critical temperature for the thermodynamic transition is reduced by the correlation effects taken into account in the Hubbard I approximation and becomes  non-monotonic. From Fig.~\ref{fig:f5} we see that the temperature of the thermodynamic transition  decreases with increasing $t_{2}$ for each $U$.

\section{Conclusion}
\label{sec:s5}

In this paper we studied correlation effects on the phase diagram of the topological insulator at finite temperatures. We started with  the definition of topological phases of density matrices and using it we extended a concept of topological invariants to nonzero temperatures. We introduced a model which allows existence of states with a long-range order, due to the interaction,  and with nontrivial topology, due to a complex hopping amplitude. 
Using  the concept of the topological Hamiltonian and the dynamical local approximation\cite{dmft1,dmft2} to many-body Hamiltonian we  found analytically the condition for the topological phase transition.
Within the Hartree and the Hubbard I approximations the complete phase diagrams with normal, topological, and long-range order phases were determined. 
The  comparison of thermodynamic phase diagrams allows us to conclude that both the temperature and the interaction play a crucial role for the existence of nontrivial topological states and they should be included in realistic calculations. 
In particular, within the Hubbard I approximations, where the correlation effects are taken into account,  the topological state  appears only inside the CDW phase and is bounded to a finite area on the phase diagram. Moreover, the homogeneous phase is a  topologically trivial Mott insulator. The Hubbard I approximation seems to be the simplest approximation that can describe  complete destruction of the topological phase only due to local correlations.

\section*{Acknowledgments}
We thank J. Tworzydlo for discussions. We acknowledge support from the Foundation for Polish Science (FNP) through the TEAM/2010-6/2 project, co-financed by the EU European Regional Development Fund.

\appendix

\section{Stability of  CDW phase}
\label{sec:a1}

 In the strong interaction limit the KMFK model is mapped onto the antiferromagnetic Ising model, where $s_{i}=n_{i}-1/2$ is a spin variable at a site $i$ and an exchange  coupling constants are $J_{1}=t_{1}^2/4U$ for the nearest-neighbors and $J_{2}=t_{2}^2/4U$ for the next-nearest-neighbors hopping, respectively.\cite{eff,eff2} 
 On the honeycomb lattice the ground state is a simple antiferromagnet for $J_{2}/J\le0.25$ and a super-antiferromagnet otherwise.\cite{ground} 
 Such spin configurations correspond  to the checkerboard and stripes pattern of the localized particles, respectively. 
 Hence, we obtain the  threshold ratio $t_{2}/t_{1}<0.5$ of the  CDW phase stability for  arbitrary U. 
 When it is not satisfied the  CDW phase can still exist but only in a finite range of U.

\section{Change of the topological invariant}
\label{sec:a2}

The topological invariant of the KMFK model (\ref{eqn:e1}) is equivalent to the Chern number for the spin up copy of the Haldane model.\cite{chern} Chern number is defined by\cite{kanerev}
\begin{equation}
\label{eqn:a2e1}
C=\frac{1}{2\pi}\int_{BZ}\Omega({\bf k})\, d{\bf k}
\end{equation}
where the surface integral is over the Brillouin zone (BZ), and  $\Omega({\bf k})=\nabla_{{\bf k}}\times A({\bf k})$ is a Berry curvature.
The vector potential  $A({\bf k})=-i\langle u({\bf k})|\nabla_{{\bf k}}|u({\bf k})\rangle$ is  called a  Berry connection and $|u({\bf k})\rangle$ are eigenvectors of the lower block  band of the Hamiltonian matrix (\ref{eqn:e5}). From Stokes  theorem we see that Chern number is  nonzero when  one global gauge of vector potential  the $A({\bf k})$ does not exist. Therefore, this observation  yields the practical test of  the  existence of  a  nontrivial topology.

In the KMFK model, the topological Hamiltonian matrix is given by two dimensional matrix which can be written in the form
\begin{equation}
\label{eqn:a2e2}
H^{\rm top}({\bf k})=H^{x}({\bf k})\hat{\tau}^{x}+H^{y}({\bf k})\hat{\tau}^{y}+H^{z}({\bf k})\hat{\tau}^{z},
\end{equation}
where the Pauli matrices are marked by $\tau^{i}$ with $i=x,y,z$ and the decomposition  coefficients $H^{i}({\bf k})$ are functions of  the momentum. The spin index $\sigma$ is omitted. 
For the two dimensional Hamiltonian matrix  the eigenvector of the lower band, i.e. the eigenvalue for a given $\bf k$,  is
\begin{equation}
\label{eqn:a2e3}
u^{I}({\bf k})=\frac{1}{N^{I}}\left(
\begin{array} {c}
H^{z}({\bf k})-|H({\bf k})|\\
H^{x}({\bf k})+iH_{y}({\bf k})
\end{array}
\right),
\end{equation}
where $|H({\bf k})|=\sqrt{H^{x}({\bf k})^{2}+H^{y}({\bf k})^{2}+H^{z}({\bf k})^{2}}$ and $N^{I}$ is a normalization constant. 
This wave function is singular when $H^{x}({\bf k})=H^{y}({\bf k})=0$ and $H^{z}({\bf k})>0$. 
We can gauge out  the singularity by performing a gauge transformation $u^{I}({\bf k})\rightarrow u^{I}({\bf k})e^{i\phi({\bf k})}$ obtaining the eigenvector 
\begin{equation}
\label{eqn:a2e4}
u^{II}({\bf k})=u^{I}({\bf k})e^{i\phi({\bf k})}=\frac{1}{N^{II}}\left(
\begin{array} {c}
-H^{x}({\bf k})+iH^{y}({\bf k})\\
H^{z}({\bf k})+|H({\bf k})|
\end{array}\right).
\end{equation}
It is now singular when $H^{x}({\bf k})=H^{y}({\bf k})=0$ and $H^{z}({\bf k})<0$. 
The vector $\vec{d}({\bf k})=(H^{x}({\bf k}), H^{y}({\bf k}), H^{z}({\bf k}))/ |H({\bf k})| $ sets a map from the BZ into the  two dimensional unit sphere.
If $\vec{d}(BZ)$ covers both poles of the sphere there is no a single gauge  in the whole BZ to avoid the  singularities.

For the  KMFK model $H^{x}=H^{y}=0$ only at the  Dirac points ${\bf K}=(4\pi/3\sqrt{3}a,0)$ and ${\bf K}^{'}=-{\bf K}$, where $a$ is a lattice constant. 
At these points $H^{z}({\bf K})=\Sigma_{R}(d)-3\sqrt{3}t_2$ and $H^{z}({\bf K}^{'})=\Sigma_{R}(d)+3\sqrt{3}t_2$. 
Hence, the vector  $\vec{d}({\bf k })$ contains only one pole for ${\bf k}\in$BZ  when the  condition
\begin{equation}
\label{eqn:ea2e5}
|\Sigma_{R}(d)|>|3\sqrt{3}t_2|
\end{equation}
is fulfilled and therefore  the system is in the topologically  trivial phase.


\begin{thebibliography}{47}
\bibitem{kane} C. L. Kane and E. J. Mele, Phys. Rev. Lett. \textbf{95}, 226801 (2005).
\bibitem{kanez2} C.L. Kane, E.J. Mele, Phys. Rev. Lett. \textbf{95}, 146802 (2005).
\bibitem{kane3d} L. Fu, C.L. Kane, E.J. Mele, Phys. Rev. Lett. \textbf{98}, 106803 (2007).
\bibitem{zhang} B. A. Bernevig, T. L. Hughes, S.-C. Zhang, Science \textbf{314}, 1757 (2006).
\bibitem{kane3d2} L. Fu, C.L. Kane Phys. Rev. B \textbf{76} 045302 (2007).
\bibitem{kanerev}M. Z. Hasan and C. L. Kane, Rev. Mod. Phys. \textbf{82}, 3045 (2010).
\bibitem{zhangrev} X. L. Qi and S. C. Zhang, Rev. Mod. Phys. \textbf{83}, 1057 (2011).
\bibitem{geo} J. C. Budich, B. Trauzettel, physica status solidi \textbf{7}, 109 (2013).
\bibitem{hg} M. Konig, S. Wiedmann, C. Brune, A. Roth, H. Buhmann, L. W. Molenkamp,X. L. Qi, and S. C. Zhang, Science \textbf{318}, 766 (2007).
\bibitem{pl} P. Dziawa, B. J. Kowalski, K. Dybko, R. Buczko, A. Szczerbakow, M. Szot, E. Lusakowska, T. Balasubramanian, B. M. Wojek, M. H. Berntsen, O. Tjernberg, T. Story, Nature Materials \textbf{11}, 1023-1027 (2012).
\bibitem{bi} H. Peng,  et al., Nat. Mater. \textbf{9}, 225–229 (2010).
\bibitem{bi2}  J. G. Checkelsky,  et al., Phys. Rev. Lett. \textbf{103}, 246601 (2009).
\bibitem{spec1} D. Hsieh, et al., Nature \textbf{452}, 970 (2008).
\bibitem{spec2} D. Hsieh,  et al., Phys. Rev. Lett. \textbf{103}, 146401 (2009).
\bibitem{spec3} Y. Xia, et al., Nat. Phys. \textbf{5}, 398 (2009).
\bibitem{corrrev} M. Hohenadler, F. F. Assaad, J. Phys.: Condens. Matter \textbf{25}, 143201 (2013).
\bibitem{corr1} S. Raghu, Xiao-Liang Qi, C. Honerkamp, S.-C. Zhang, Phys.Rev.Lett. \textbf{100}, 156401 (2008).
\bibitem{corr2} L. Wang, et al., arXiv:1012.5163 (2010).
\bibitem{corr3} J. C. Budich, B. Trauzettel, G. Sangiovanni, Phys. Rev. B \textbf{87}, 235104 (2013).
\bibitem{top1} A. Shitade, et al., Phys. Rev. Lett. \textbf{102}, 256403 (2009).
\bibitem{opthal} G.Jotzu, et. al, Nature \textbf{515}, 237-240 (2014).
\bibitem{opt} I. Bloch, J. Dalibard, W. Zwerger,Rev. Mod. Phys. \textbf{80}, 885 (2008).
\bibitem{spielmann} V. Galitski and I.B. Spielman, Nature {\bf 494}, 49 (2013).
\bibitem{term1} Y.-X. Zhu, et al., J. Phys.: Condens. Matter \textbf{26}, 175601 (2014).
\bibitem{term2} Y.-H. Chen, et al., Phys. Rev. B \textbf{91}, 045122 (2015).
\bibitem{term3} T. Yoshida, S. Fujimoto, N. Kawakami, Phys. Rev. B. \textbf{85}, 125113 (2012).
\bibitem{term4} W. Wu, et al., Phys. Rev. B \textbf{85}, 205102 (2012).
\bibitem{topden} J.C. Budich, S. Diehl, Phys. Rev. B \textbf{91}, 165140 (2015)
\bibitem{falicov}L. M. Falicov and J. C. Kimball, Phys. Rev. Lett. \textbf{22}, 997 (1969).
\bibitem{halfal} H.-S. Nguyen, M.-T. Tran, Phys. Rev. B \textbf{88}, 165132 (2013).
\bibitem{dmft1} W. Metzner, D. Vollhardt, Phys. Rev. Lett. \textbf{62}, 324 (1989).
\bibitem{dmft2} A. Georges, G. Kotliar, W. Krauth, and M. J. Rozenberg, Rev. Mod. Phys. \textbf{68}, 13 (1996).
\bibitem{topclass}  A. P. Schnyder, S.i Ryu, A. Furusaki, A. W. W. Ludwig, Phys. Rev. B \textbf{78}, 195125 (2008)
\bibitem{noneq1} J.C. Budich, P. Zoller, S. Diehl 	Phys. Rev. A \textbf{91}, 042117 (2015)
\bibitem{noneq2} S. Diehl, E. Rico, M. A. Baranov, P. Zoller Nature Physics \textbf{7}, 971 (2011)
\bibitem{intcon} Z. Wang, X.-L. Qi, S.-C. Zhang, Phys.Rev.Lett. \textbf{105},256803 (2010)
\bibitem{inv1} Z. Huang, D. P. Arovas, Phys. Rev. Lett. \textbf{113}, 076407 (2014).
\bibitem{inv2} O. Viyuela, A. Rivas, M. A. Martin-Delgado, Phys. Rev.Lett. \textbf{113}, 076408 (2014).

\bibitem{hirsch} J.E. Hirsch, Phys. Rev. Lett. {\bf 83}, 1834 (1999).
\bibitem{haldane}  F. D. M. Haldane, Phys. Rev. Lett. \textbf{61}, 2015 (1988).
\bibitem{lieb}  T. Kennedy, E. H. Lieb,  Physica A \textbf{138}, 320 (1986).
\bibitem{wojt} J. Wojtkiewicz, Journal of Statistical Physics \textbf{123}, 585-600 (2006).

\bibitem{chern} For Haldane model Chern number $C=\pm1$ depends on the sign of the Peierls phase.\cite{haldane} In the case of the system from Eq. (\ref{eqn:e1}), the total Chern number is sum of Chern numbers for both copies of the Haldane model  $C=C_{\uparrow}+C_{\downarrow}=0$. However, we can define a new topological invariant connected with conservation of the perpendicular spin $S_{z}$. It is called the spin Chern number \cite{kane} and given by the formula $C_{s}=\frac{C_{\uparrow}-C_{\downarrow}}{2}$. Hence, for our model it is equal to the Chern number of the single copy of the Haldane model $C_{s}=C_{\uparrow}$. The topological invariant $\upsilon$ of time reversal symmetry is linked to spin Chern number by the equation $\upsilon=C_{s}\bmod2$.\cite{kanerev}

\bibitem{topham1} Z. Wang and S.-C. Zhang,  Phys. Rev. X \textbf{2}, 031008 (2012).
\bibitem{topham2} Z. Wang and B. Yan, J. Phys.: Condens. Matter \textbf{25}, 15560 (2013).
\bibitem{booktop} B. A. Bernevig and T. L. Hughes, Topological Insulators and Topological Superconductors, Princeton University Press, pp. 30-32 (2013).
\bibitem{bookgreen} F. Gebhard, The Mott Metal-insulator Transition: Models and Methods, Springer, pp. 270-274 (1997).
\bibitem{freq} L. Wang, Xi Dai, X. C. Xie, Phys. Rev. B \textbf{84}, 205116 (2011).
\bibitem{Lutt} T. D. Stanescu, P. W. Phillips, T.-P. Choy, Physical Review B, \textbf{75}, 104503 (2007).
\bibitem{future} D. Zdulski and K. Byczuk, in preparation.
\bibitem{free} A.M. Shvaika, J.K. Freericks, Phys. Rev. B \textbf{67}, 153103 (2003).
\bibitem{next} N. Varney, K. Sun, M. Rigol, and V. Galitski, Phys. Rev. B \textbf{84}, 241105(R) (2011).
\bibitem{eff}  Umesh K. Yadav, T. Maitra, Ishwar Singh, arXiv:1205.6555 (2012).
\bibitem{eff2} Ch. Gruber, N. Macris, A. Messager, D. Ueltschi, Journal of Statistical Physics, vol. \textbf{86}, pp. 57-108 (1997).
\bibitem{ground} A. O'Hare, F.V. Kusmartsev, K.I. Kugel, International Journal of Modern Physics B, \textbf{23} (20-21), pp. 3951-3967 (2009).













\end{thebibliography}
\end{document}